\documentstyle[prc,multicol,epsf,aps]{revtex}
\begin{document} 
\draft
\tighten 
\title{On the uncertainty in the $0\nu\beta\beta$ decay
nuclear matrix elements}

\author{V. A. Rodin$^1$, Amand Faessler$^1$, F. \v{S}imkovic$^2$, 
and Petr Vogel$^{1,3,4}$ }

\address{$^1$Institute f\"{u}r Theoretische Physik der Universit\"{a}t
T\"{u}bingen, D-72076 T\"{u}bingen, Germany}

\address{$^2$Department of Nuclear Physics, Comenius University,
Bratislava, Slovakia} 

\address{$^3$Department of Physics 106-38, California Institute
of Technology, Pasadena, CA 91125, USA}

\address{$^4$Institute of Particle and  Nuclear Physics, 
Charles University, Prague, Czech Republic}

\date{\today} 

\maketitle

\begin{abstract}
The nuclear matrix elements $M^{0\nu}$  of the
neutrinoless double beta decay ($0\nu\beta\beta$) are evaluated for
$^{76}$Ge,$^{100}$Mo,$^{130}$Te, and $^{136}$Xe within the Renormalized
Quasiparticle Random Phase Approximation (RQRPA) and the simple QRPA.
Three sets of single particle level schemes are used, ranging in size from 9
to 23 orbits. When the strength of the
particle-particle interaction is adjusted
so that the $2\nu\beta\beta$
decay rate is correctly reproduced,  
the resulting
$M^{0\nu}$ values become essentially independent on the size of the basis, and
on the form of
different realistic nucleon-nucleon potentials. Thus, one of the
main reasons for variability of the calculated $M^{0\nu}$ 
within these methods is eliminated. 
\end{abstract} 
\pacs{ 23.10.-s, 21.60.-n, 23.40.Hc} 

\begin{multicols}{2}


The discovery of oscillations of atmospheric \cite{SK}, solar \cite{SNO},
and reactor \cite{KAM} neutrinos shows that neutrinos have a non-vanishing
rest-mass. However, since the study of oscillations provides only
information on the mass-squared differences, it
cannot by itself determine the absolute values of the masses (or even,
at present, the sign of the mass-squared differences). Moreover, flavor
oscillations are insensitive to the charge conjugation properties of the
neutrinos, i.e. whether massive neutrinos are Dirac or Majorana particles.
Study of the neutrinoless double beta decay ($0\nu\beta\beta$) is the best
potential source of this crucial information on the Majorana nature of
the neutrinos and on the absolute mass scale.

The search for $0\nu\beta\beta$ decay has a long history. So far, the decay
has not been seen, but impressive limits on its half-life have been established
(for the review of the field, see e.g. \cite{EV2002} or \cite{FS1998}).
With the progress in oscillation studies, which now established the
mass scale $m_{scale} \approx \sqrt{\Delta m^2}$, a wave of enthusiasm emerged
in the community to develop a new generation of experiments, sensitive
to such mass scale.  In that context it is crucial to develop 
in parallel methods
capable of reliably evaluating the nuclear matrix elements, and realistically
asses their uncertainties. The goal of the present work is to make
a contribution to that endeavor. 

Clearly, the determination of the effective Majorana mass 
$\langle m_{\nu} \rangle$ can be only as good as the knowledge of
the nuclear matrix elements, since the half-life of the $0\nu\beta\beta$
decay and $\langle m_{\nu} \rangle$ are related by 
\begin{equation}
\frac{1}{T_{1/2}} = G^{0\nu}(E_0,Z) |M^{0\nu}|^2 |\langle m_{\nu} \rangle|^2~,
\end{equation}
where $G^{0\nu}(E_0,Z)$ is the precisely calculable phase-space factor,
and $M^{0\nu}$ is the corresponding nuclear matrix element. 
Thus, obviously, any uncertainty in $M^{0\nu}$ makes the value
of $\langle m_{\nu} \rangle$ equally uncertain. In turn,
\begin{equation}
\langle m_{\nu} \rangle = \sum_i^N |U_{ei}|^2 e^{i\alpha_i} m_i ~,
~({\rm all~} m_i \ge 0)~,
\end{equation}
where $\alpha_i$ are the unknown Majorana phases. The elements of the
mixing matrix $|U_{ei}|^2$ and the mass-squared differences $\Delta m^2$
can be determined in the oscillation experiments. Using the present
knowledge of these quantities, (see, e.g., \cite{PP2002}) 
and limits on $T_{1/2}$, one could
decide whether the neutrino mass pattern is degenerate,
or follows the inverse or normal hierarchies. If, on the other hand,
the existence of the $0\nu\beta\beta$ decay is proven and the 
value of $T_{1/2}$ is found, a relatively narrow range of
absolute neutrino mass scale can be determined, independently of the
phases $\alpha_i$ in most situations \cite{EV2002,PP2002}. However,
such an important insight is possible only if the nuclear matrix elements
are accurately known. 
 
The nuclear matrix element $M^{0\nu}$ is defined as
\begin{equation}
M^{0\nu} = \langle f | -\frac{M^{0\nu}_F}{g^2_A} + M^{0\nu}_{GT} + 
M^{0\nu}_T |i\rangle
\end{equation}
where $|i\rangle, ~(|f\rangle )$ are the wave functions of the ground
states of the initial (final) nuclei.  The
explicit forms of the operators $M^{0\nu}_F, M^{0\nu}_{GT}$ and
$M^{0\nu}_T$ are given in Ref. \cite{si99}.
In comparison
with most of previous $0\nu\beta\beta$ decay studies 
\cite{SSPF1997,VZ86,CFT87,MK88} the higher 
order terms of the nucleon current are also included
in the present calculation, resulting 
in a suppression of the nuclear matrix element by about 30 \% \cite{si99}.
Note that in the numerical calculation of $M^{0\nu}$ here the closure 
approximation is avoided and the unquenched values $g_A = 1.25$, $g_V = 1.0$
are used.

Two basic methods are used in the evaluation of $M^{0\nu}$, 
the quasiparticle random phase approximation (QRPA)
with its various modifications and the nuclear
shell model (NSM). These two approaches represent in some
sense opposite extremes.

In the QRPA one can include essentially unlimited set 
of single-particle states, but only
a limited subset of configurations (iterations of the particle-hole,
respectively two-quasiparticle configurations). 
In the context of QRPA several issues that have been raised,
and deserve a systematic study:
\begin{itemize}
\item For realistic interactions the QRPA solutions are near the point
of the so-called collapse and thus its applicability is questionable.
Numerous attempts have been made to extend the method's range of validity
by partially avoiding the violation of the
Pauli principle. Here we consider the simplest of them,
the renormalized QRPA (RQRPA) \cite{TS1995,SSPF1997}. We compare the
results of RQRPA with those of the standard QRPA 
\cite{VZ86,CFT87,MK88}.
\item The choice of the size of 
single-particle (s.p.) space is to some extent arbitrary,
often dictated by convenience. What effect this choice has on the
$M^{0\nu}$ values is the main thrust of the present work.
\item There are various forms of the nucleon-nucleon potential that
lead to somewhat different forms of the resulting $G$-matrix
\footnote{Brueckner reaction matrix elements with these
nucleon-nucleon potentials were calculated
and used in Refs. \cite{Muether1,Muether2,Muether3}.}. By comparing 
the results obtained with three such potentials (Bonn-CD \cite{Bonn}, 
Argonne \cite{Argonne}, and Nijmegen \cite{Nijmegen})
we show that the
resulting $M^{0\nu}$ are essentially identical, and independent
of the choice of the realistic nucleon-nucleon potential.
\end{itemize}

In contrast, in the NSM one chooses a limited set of
single-particle states in the vicinity of the Fermi level, 
and includes all (or most) configurations of
the valence nucleons on these orbits in the evaluation of $M^{0\nu}$.
The main open question in this approach is to determine 
the effects of the neglected single-particle
states further away from the Fermi level.
As shown below, we have also performed QRPA and RQRPA calculations
with the set of s.p. states used usually in NSM.
It appears that these methods, at least with the nucleon-nucleon
interaction we used, are not applicable for such small s.p.
bases. 

In what follows the $0\nu\beta\beta$ decay nuclear 
matrix elements $M^{0\nu}$ 
for $^{76}$Ge, $^{100}$Mo, $^{130}$Te, and $^{136}$Xe are evaluated.
These nuclei are most often considered as candidate sources for the
next generation of the experimental search for $0\nu\beta\beta$ decay.
For each of them three choices of the s.p. basis are
 considered.
The smallest one has 9 levels (oscillator shells $N$=3,4) for $^{76}$Ge,
and 13 levels (oscillator shells $N$=3,4 plus the $f + h$ orbits from $N=5$)
for $^{100}$Mo,  $^{130}$Te and $^{136}$Xe. For the intermediate size s.p. base
the $N=2$ shell in $^{76}$Ge and $^{100}$Mo are added, and 
for  $^{130}$Te and $^{136}$Xe  also  the $p$ orbits from $N = 5$. 
Finally, the largest
s.p. space \cite{SSPF1997} contains 21 levels for $^{76}$Ge and $^{100}$Mo
(all states from shells $N=1-5$), and 23 levels for $^{130}$Te and $^{136}$Xe
($N=1-5$ and $i$ orbits from $N=6$). Thus the smallest set corresponds
to $1 \hbar\omega$ particle-hole excitations, and the largest
to about $4 \hbar\omega$ excitations. The s.p. energies have been calculated
with the Coulomb corrected Woods-Saxon potential.

It is well known that the residual interaction  
is an effective interaction that depends on the size of the
single-particle basis. Hence, when the basis is changed, the interaction
should be modified as well. Here we propose a rather simple way to
accomplish the needed renormalization. 

There are three important ingredients in QRPA and RQRPA. First, the pairing
interaction has to be included 
by solving the corresponding gap equations. Within the BCS
method the strength of the pairing interaction depends on the size
of the s.p. basis. As usual, we multiply
the pairing part of the interaction by a factor $g_{pair}$
whose magnitude is adjusted, for both protons and neutrons, such
that the pairing gap is correctly reproduced, separately for the
initial and final nuclei. 

Second,  QRPA equations of motion contain a block corresponding to
the particle-hole interaction, renormalized by an overall strength
parameter $g_{ph}$. That parameter is typically adjusted by requiring
that the energy of some chosen collective state, often the giant GT
resonance, is correctly reproduced. We find that the calculated energy 
of the giant GT state is almost independent of the size of the s.p.
basis and is well reproduced with $g_{ph} \approx 1$. Accordingly, we use
$g_{ph} = 1$ throughout, without adjustment.

Finally, QRPA equations of motion contain a block corresponding to
the particle-particle interaction, renormalized by an overall strength 
parameter $g_{pp}$. (The importance of the particle-particle interaction
for the $\beta$ strength was recognized first in Ref. \cite{Cha83}, and
for the $\beta\beta$ decay in \cite{VZ86}.) It is well known that the
decay rate for both modes of $\beta\beta$ decay depends sensitively
on the value of $g_{pp}$. In the following we use this property to
find the value of $g_{pp}$ for each of the possible s.p. bases.
The value of the parameter $g_{pp}$ is fixed 
in each case so that the known half-life of the 
$2\nu \beta\beta$ decay is correctly reproduced. 
The $2\nu\beta\beta$ half-lives and average matrix elements
collected in Table 1 of Ref. \cite{EV2002} are used, where the original 
references to the corresponding experiments can be found. A similar
compilation of the  $2\nu\beta\beta$ data can be found in Ref. \cite{Bar02}.
The resulting
adjusted values of $g_{pp}$ are shown in Table \ref{tab:gpp}
for both the RQRPA and QRPA methods. One can see
that as the basis increases, the effective $g_{pp}$ decreases
\footnote{For $^{136}$Xe we use the lower limit on the $2\nu$ half-life
for the adjustment of $g_{pp}$; hence
further adjustment might 
be needed when the $2\nu$ half-life becomes known.},
as expected.

The adjustment of $g_{pp}$
is a crucial point of the present work. Several studies of the
sensitivity of  the nuclear matrix elements $M^{0\nu}$ to various
modifications of the QRPA method as well as to the number of s.p.
states and values of $g_{pp}$ were made in the recent past
\cite{SSPF1997,SK01,CS03}. Typically, these studies concluded that
the values of $M^{0\nu}$  vary substantially depending on
all of these things. Here we come to a different conclusion: by requiring
that the known $2\nu\beta\beta$ decay half-life is correctly reproduced,
and adjusting the parameter $g_{pp}$ accordingly, we remove much
of the sensivity on the number of single-particle states, on the 
$NN$ potential employed, and even on whether RQRPA or just simple
QRPA methods are used. 
  
Clearly, the chosen procedure of finding the effective interaction
is rather crude. Ideally, properly evaluated effective Hamiltonian,
as well as the corresponding effective $\beta\beta$ operator should
be used in each case. However, as shown below, the chosen
procedure appears to be  sufficient to stabilize
the values of the nuclear matrix elements $M^{0\nu}$.
Note, that we did not use the quenched value of the axial
current coupling constant $g_A$, as is often done in studies
of ordinary $\beta$ decay. We are convinced that with the
quenched  $g_A$ the basic conclusion of our work will be 
similar, even though the values of $g_{pp}$ in Table \ref{tab:gpp}
would be of course different.

Having fixed the parameters of the effective Hamiltonian we can proceed
and evaluate the $0\nu\beta\beta$ nuclear matrix elements $M^{0\nu}$
and then the corresponding half-life (we list the phase-space factors
in Table \ref{tab:amp} in units of $10^{-25}$ years 
for $\langle m_{\nu} \rangle$ in eV.)
In evaluating $M^{0\nu}$ we take the short-range correlations into
account in the standard way, i.e. by multiplying the operators with
the square of the correlation function \cite{MS1976}
\begin{equation}
f(r) = 1 - e^{-ar^2}(1 - br^2); ~a = 1.1 ~{\rm fm}^{-2},
~b = 0.68 ~{\rm fm}^{-2}.
\end{equation}
Note that the effect of short-range correlations reduces the
matrix elements $M^{0\nu}$ by a factor of about two,
in agreement with other evaluations.

\end{multicols}
\widetext

\begin{table}[htb]
  \begin{center}
    \squeezetable
    \caption{Values of the effective particle-particle strength
parameter $g_{pp}$ for the
different nuclei and different sizes of the single-particle space. The
values in col. 3 correspond to the minimal s.p. space, in column 4
to the intermediate one, and in col. 5 to the largest s.p. space.
In each of these there are three entries
corresponding to the G-matrix based on the Bonn, Argonne,
and Nijmegen potentials (in that order). 
For every considered nucleus the upper line was obtained
in RQRPA and the lower one in the simple QRPA.
In column 2 we give the
corresponding half-life of the $2\nu \beta\beta$ decay used in the
adjustment (see text, for $^{136}$Xe the lower limit is used).}
    \label{tab:gpp}
    \renewcommand{\arraystretch}{1.2}
    \begin{tabular}{|c|c|c|c|c|}
  Nucleus &  $T_{1/2}^{2\nu}$ (in $10^{20}$ y) &
min. s.p. space & interm. s.p. space
& largest s.p. space \\
\hline \hline
$^{76}$Ge & 13.0 & 0.99 1.12 1.07  & 0.88 1.00 0.95 & 0.79 0.88 0.84 \\
 & & 0.85 0.96 0.91  & 0.75 0.85 0.81 & 0.65 0.72 0.69 \\
\hline
$^{100}$Mo & 0.08 & 1.21 1.35 1.30 & 1.09 1.21 1.17 & 1.00 1.10 1.07 \\
 & &  0.94 1.03 1.00 & 0.85 0.93 0.90 & 0.78 0.85 0.83 \\
\hline
$^{130}$Te  & 27.0 & 0.97 1.10 1.05 & 0.90 1.01 0.97 & 0.84 0.94 0.90 \\
 & & 0.84 0.94 0.90 & 0.78 0.86 0.83 & 0.71 0.78 0.76 \\
\hline
$^{136}$Xe  & 8.1 & 0.82 0.93 0.89 & 0.77 0.87 0.83 & 0.72 0.82 0.78 \\
 & & 0.73 0.83 0.79 & 0.68 0.77 0.74 & 0.62 0.70 0.67
  \end{tabular}
  \end{center}
\end{table}


\begin{table}[htb]
  \begin{center}
    \squeezetable
    \caption{Values of the nuclear matrix element $|M^{0\nu}|$ for the
different nuclei and different sizes of the single-particle space. 
In column 2 is
corresponding phase space factor $G^{0\nu}$.
For the explanation of the notation in columns 3-5 see the caption of Table 
\protect{\ref{tab:gpp}}.}
    \label{tab:amp}
    \renewcommand{\arraystretch}{1.2}
    \begin{tabular}{|c|c|c|c|c|}
  Nucleus &  $G^{0\nu}$ (in $10^{-25}$y$\cdot$eV$^{-2}$ &
min. s.p. space & interm. s.p. space
& largest s.p. space \\
\hline \hline
$^{76}$Ge & 0.30 & 2.41 2.37 2.35  & 2.52 2.44 2.47 & 2.32 2.34 2.35  \\
  & & 2.68 2.62 2.65 & 2.81 2.74 2.72 & 2.62 2.64 2.64 \\
\hline
$^{100}$Mo & 2.19 & 1.08 1.08 1.05 & 1.12 1.14 1.08 & 1.28 1.34 1.27 \\
  & & 1.19 1.22 1.19 & 1.25 1.28 1.20 & 1.38 1.41 1.40 \\
\hline
$^{130}$Te  & 2.12 & 1.42 1.32 1.34 & 1.40 1.33 1.32 & 1.17 1.13 1.14 \\
  & &  1.50 1.42 1.40 & 1.46 1.42 1.39 & 1.22 1.17 1.15 \\
\hline
$^{136}$Xe  & 2.26 & 1.05 0.99 0.99 & 1.09 1.03 1.04 & 0.92 0.83 0.86 \\
  & & 1.11 1.02 1.05 & 1.15 1.06 1.06 & 0.98 0.90 0.91
  \end{tabular}
  \end{center}
\end{table}

\begin{multicols}{2}

As pointed out earlier, in the nuclear shell model an even smaller
set of single-particle states is used corresponding
to $0\hbar\omega$. This choice reflects
the practical computational limitations in handling
the extremely large number of possible configurations,
while it seems to be sufficient to describe the spectroscopy
of low-lying nuclear states. In the NSM evaluation of the $\beta\beta$
decay rates \cite{SM96}  four s.p. orbits
($f_{5/2}, p_{3/2}, p_{1/2},g_{9/2}$) were used for $^{76}$Ge and
a five orbits ($d_{5/2}, d_{3/2}, s_{1/2}, g_{7/2}, h_{11/2}$)
for $^{130}$Te and $^{136}$Xe. These s.p. sets are free
of the spurious center-of-mass states, but obviously miss a large
part of the GT strength as well as of the strength corresponding
to the higher multipoles. In order to describe GT transitions between
low-lying states in the NSM, it is necessary to quench the corresponding
strength. This is most conveniently formally achieved by using $g_A = 1.0$
instead of the free nucleon value of $g_A = 1.25$. We follow this prescription
in our attempt to use this smallest s.p. space, and only there.

It appears that it is impossible to describe the $2\nu\beta\beta$ 
decay in such
s.p. space using QRPA or RQRPA, and the nucleon-nucleon potentials
employed in this work. One would have to renormalize the particle-particle 
block too much, with $g_{pp} \sim 2.0$, unlike the rather modest 
renormalization shown in Table \ref{tab:gpp}. With such large value of
$g_{pp}$ the interaction is too far removed from the $G$-matrix used in
the rest of this work. Therefore, one cannot expect to obtain
sensible  $0\nu$ matrix elements. In fact, we obtained very small
matrix elements in this case for  $^{130}$Te and $^{136}$Xe, while,
perhaps accidentally, for $^{76}$Ge they are in
a crude agreement with the NSM result \cite{SM96}.       

\vspace{0.2cm}

\begin{figure}[htb]
  \begin{center}
    \leavevmode
    \epsfxsize=0.4\textwidth
    \epsffile{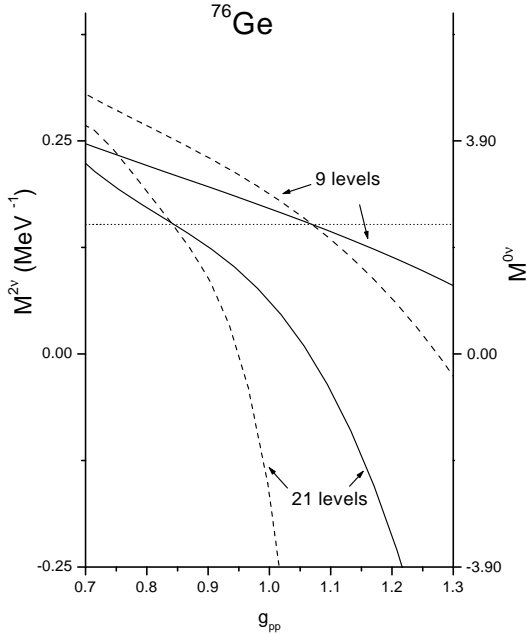}
    \caption{Dependence of the matrix elements $M_{2\nu\beta\beta}$
    (left scale, dashed lines) and $M_{0\nu\beta\beta}$
     (right scale, full lines) on the parameter $g_{pp}$. Calculations
were performed for 9 and 21 s.p. levels for $^{76}$Ge as indicated;
the Nijmegen potential and RQRPA method were used. The thin dotted 
horizontal line indicates that by fixing $g_{pp}$ to reproduce the
experimental value $M_{2\nu\beta\beta} = 0.15$ MeV$^{-1}$ the
value of $M_{0\nu\beta\beta}$ is also stabilized. }
    \label{fig:Ge}
  \end{center}
\end{figure}

We list the results with the three larger single-particle
bases in Table \ref{tab:amp} which represents the
most significant part of the present work. As one can see by inspecting
the entries, one can draw two important conclusions:
\begin{itemize}
\item The resulting $M^{0\nu}$ do not depend noticeably on the
form of the nucleon-nucleon potential used. That is not an unexpected
result.
\item Even more importantly, with our choice of $g_{pp}$ the results
are also essentially independent on the size of the s.p. basis.
This is a much less obvious and rather pleasing conclusion.
It can be contrasted with the result one would get for a constant
$g_{pp}$ independent on the size of the s.p. basis. The values of $M^{0\nu}$
differ then between the small and large bases by a factor of two or more.
\end{itemize}

The effect of the $g_{pp}$ adjustment is illustrated in Fig. \ref{fig:Ge},
showing that our procedure leads to almost constant
 $M^{0\nu}$ matrix elements. On the other hand, by choosing a fixed
value of $g_{pp}$ the resulting  $M^{0\nu}$ matrix elements for
9 and 21 s.p. levels would differ substantially.

\begin{figure}[htb]
  \begin{center}
    \leavevmode
    \epsfxsize=0.4\textwidth
    \epsffile{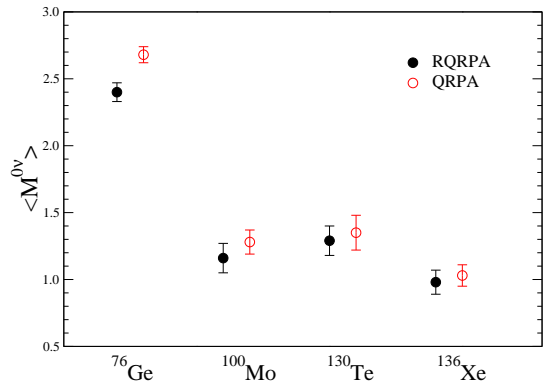}
    \caption{Average nuclear matrix elements 
$\langle M^{0\nu} \rangle $
and their variance for both methods and for the four considered nuclei.}
    \label{fig:Fedor}
  \end{center}
\end{figure}

The entries in Table \ref{tab:amp} are relatively close to each other.
To emphasize this feature, each calculated value is treated
as an independent determination and
for each nucleus the corresponding
average  $\langle M^{0\nu} \rangle$ 
matrix elements (averaged over the three potentials
and the three choices of the s.p. space) is evaluated, as well as its variance
$\sigma$
\begin{equation}
\sigma^2 = \frac{1}{N - 1} 
\sum_{i=1}^N (M_i^{0\nu} - \langle M^{0\nu} \rangle)^2, ~~(N = 9).
\end{equation}
These quantities (with 
the value of $\sigma$ in paretheses) are shown in Table \ref{tab:t12}. 
Not only is the variance substantially less than the average
value, but the results of QRPA, albeit slightly larger, are quite
close to the RQRPA values. The averaged nuclear matrix elements for
both methods and their variance are shown in Fig. \ref{fig:Fedor}.

Combining the average  $\langle M^{0\nu} \rangle$ with the phase-space
factors listed in Table \ref{tab:amp} the expected half-lives
(for RQRPA and $\langle m_{\nu} \rangle$ = 50 meV, the scale
of neutrino masses suggested by oscillation experiments) are also
shown in Table \ref{tab:t12}.
These predicted
half-lives are a bit longer (particularly for the last three nuclei on our
list) then various QRPA calculations usually predict. They are faster,
however, then the shell model results of Ref. \cite{SM96}.

\begin{table}[htb]
  \begin{center}
    \squeezetable
    \caption{Averaged $0\nu\beta\beta$ nuclear matrix elements 
 $\langle M^{0\nu} \rangle$
and their variance $\sigma$  (in parentheses)
evaluated in the RQRPA and QRPA.
In column 4 the  $0\nu\beta\beta$ half-lives
evaluated with the RQRPA average nuclear matrix element and for the
$\langle m_{\nu} \rangle$ = 50 meV are shown.}
    \label{tab:t12}
    \renewcommand{\arraystretch}{1.2}
    \begin{tabular}{|c|c|c|c|}
  Nucleus & RQRPA & QRPA & $T_{1/2}$ (in $10^{27}$ y for
$\langle m_{\nu} \rangle$ = 50 meV \\
\hline \hline
$^{76}$Ge & 2.40(0.07) & 2.68(0.06) & 2.3 \\
\hline
$^{100}$Mo & 1.16(0.11) & 1.28(0.09) & 1.4 \\
\hline
$^{130}$Te & 1.29(0.11) & 1.35(0.13) & 1.1 \\
\hline
$^{136}$Xe & 0.98(0.09) & 1.03(0.08) & 1.9
\end{tabular}
  \end{center}
\end{table}

\vspace{0.2cm}

Given the average nuclear matrix elements in Table \ref{tab:t12}
and the phase space factors in Table \ref{tab:amp} one can find a limit
(or actual value) of the effective neutrino mass $\langle m_{\nu} \rangle$
from any limit (or value) of $T_{1/2}$ from

\begin{equation}
\langle m_{\nu} \rangle ~{\rm (eV)}
= \left( T_{1/2} \times G^{0\nu} \right)^{-1/2}
\times \frac{1}{ \langle M^{0\nu} \rangle } ~.
\end{equation}

In conclusion, we have developed a ``practical'' way of stabilizing
the values of the $0\nu\beta\beta$ nuclear matrix elements against
their variation caused by the modification of the nucleon-nucleon
interaction
potential and the chosen size of the single-particle space. We have
also shown that the procedure yields very similar matrix elements for
the QRPA and RQRPA variants of the basic method. Even though we
cannot guarrantee that this basic method is trustworthy, we have eliminated,
or at least greatly reduced, the arbitrariness commonly present in
the published calculations.

\acknowledgments

We would like to thank Drs. H. Muether, A. Polls, and P. Czerski
for providing us with the Brueckner reaction matrix elements used
in the present work. V. R. was financed by the
Landesforschungsschwerpunktsprogramm Baden-Wuerttemberg
"Low Energy Neutrino Physics".
The work of F. \v{S}. was supported in part by the Deutsche
Forschungsgemeinschaft (436 SLK 17/298) and by the
VEGA Grant agency of the Slovac Republic under contract
No. 1/0249/03. The work of P. V. was  supported in part
by the "Internationales Graduiertenkolleg" GRK683/2
of the Deutsche Forschungsgemeinschaft and
by the Center for Particle Physics, project No. LN00A 006
of the Ministry of Education of the Czech Republic.

\end{multicols}

\end{document}